\documentclass[11pt,a4paper]{article}
\usepackage{jinstpub}
\usepackage{xspace}

\newcommand{\cm}{\ensuremath{\mathrm{\,cm}}\xspace}
\newcommand{\s}{\ensuremath{\mathrm{\,s}}\xspace}
\newcommand{\ns}{\ensuremath{\mathrm{\,ns}}\xspace}
\newcommand{\GeV}{\ensuremath{\mathrm{\,GeV}}\xspace}
\newcommand{\pt}{\ensuremath{p_{\mathrm{T}}}\xspace}
\newcommand{\lumi}{\ensuremath{\mathrm{2 \times 10^{34}\cm^{-2}\s^{-1}}}\xspace}
\newcommand{\hllumi}{\ensuremath{\mathrm{7.5 \times 10^{34}\cm^{-2}\s^{-1}}}\xspace}
\newcommand{\kHz}{\ensuremath{\mathrm{\,kHz}}\xspace}
\newcommand{\MHz}{\ensuremath{\mathrm{\,MHz}}\xspace}
\newcommand{\Gbps}{\ensuremath{\mathrm{\,Gbps}}\xspace}

\title{A CMS Level-1 Track Finder for the HL-LHC}
\author[1]{B. R. Yates for the CMS collaboration}
\affiliation[1]{The Ohio State University,\\
281 W Lane Ave, Columbus, OH 43210 USA}
\emailAdd{brent.yates@cern.ch}
\abstract{The High-Luminosity LHC will put significant demands on trigger systems. To control trigger thresholds, the CMS Collaboration is designing a novel Level-1 track trigger. The Outer Tracker will use modules with pairs of sensor layers to read out hits compatible with charged particles above 2-3\GeV. The system will combine these front-end trigger primitives to reconstruct tracks, providing a measurement of $\pt$, $\eta$, $\phi$, and $z_0$. This proceeding will introduce the CMS L1 track finding system: the algorithm and its estimated performance, hardware prototypes, and the unique challenges associated with this system.}
\begin{document}
\maketitle

\section{Introduction}
\begin{figure}[!htb]
    \centering
    \includegraphics[width=0.7\textwidth]{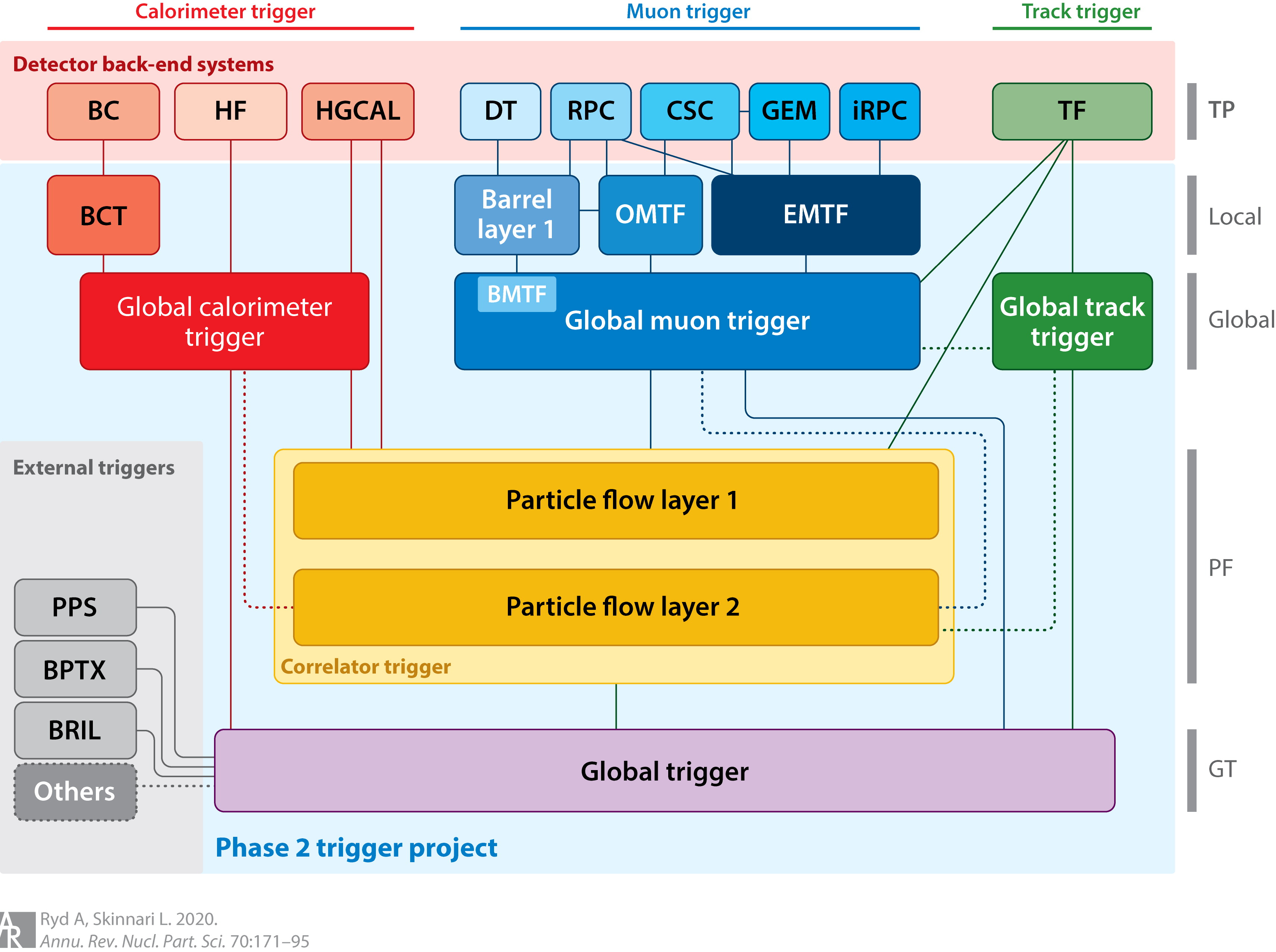}
    \caption{The full CMS L1 trigger path, including the Track Trigger, for the HL-LHC upgrade.}
    \label{fig:trigger}
\end{figure}

The increased luminosity in the proposed High Luminosity Large Hadron Collider (HL-LHC) upgrade~\cite{Zimmermann:2009zz,Apollinari:2017lan} offers some unique challenges to the continued operation of the Compact Muon Solenoid (CMS) detector~\cite{CMS:2008xjf}. With the current conditions of the LHC~\cite{Bruning:2004ej,Evans:2008zzb} a proton-proton bunch crossing (bx) occurs every 25\ns with a peak instantaneous luminosity of \lumi, resulting in a pile-up (PU) of approximately 20. The bx interval results in an initial data rate of 40\MHz, which is far too much information to read out from the detector and store on current hardware. 
The CMS detector employs a technique known as a trigger~\cite{CMS:2016ngn}. The CMS trigger consists of two main parts: the Level-1 (L1) trigger, and the High Level Trigger (HLT). The L1 trigger uses information from the calorimeters and muon system. This reduces the data rate to 100\kHz, which is then passed to the HLT server farm. The final output of the HLT is 1\kHz, which can be comfortably saved to disk for offline processing by analyzers.\\
\\
The HL-LHC upgrade will increase the instantaneous luminosity to \hllumi, increasing the PU to roughly 200. In order to handle this increased luminosity, by reducing the data rate to a reasonable value, the CMS collaboration has decided to introduce a L1 Track Trigger~\cite{Contardo:2020886}. The L1 Track Trigger will allow the CMS trigger system to include information from the precision silicon tracker, while keeping the trigger decision time at a resonable rate. The full trigger path is shown in Fig~\ref{fig:trigger}.
\\
\\
This proceeding will detail the creation of the CMS L1 Track Trigger, including: the algorithm, firmware development, target hardware, and the current status as of this publication.

\section{The L1 Track Trigger algorithm}
\begin{figure}[!htb]
    \centering
    \includegraphics[width=0.7\textwidth]{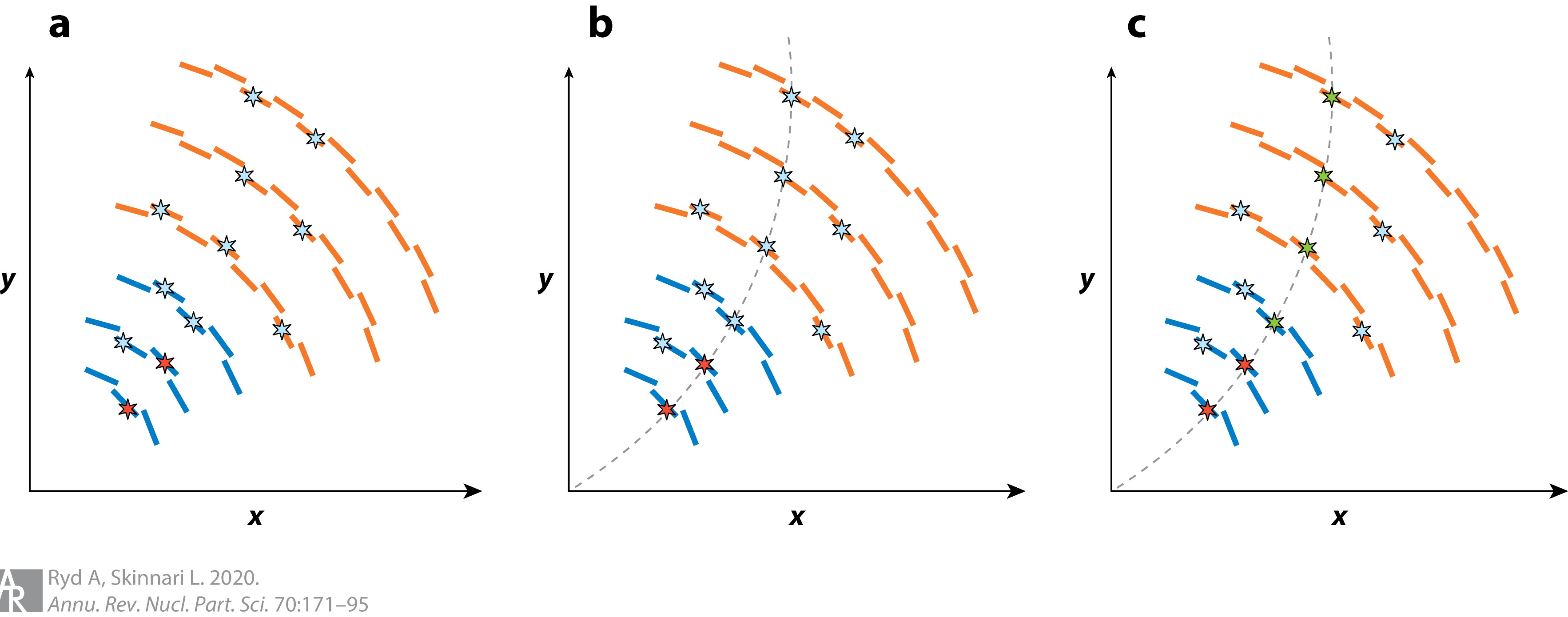}
    \caption{A diagram of the L1 Track Trigger algorithm, including the seeding (a), projecting (b), and fitting (c) stages.}
    \label{fig:algorithm}
\end{figure}

The Level-1 Track Trigger upgrade will make use of the proposed upgrades to the CMS silicon tracker. These changes include six barrel layers and five endcap disks, and a tilted geometry for the first three barrel layers. The first three layers of the barrel, and the bottom half of the endcap disks, are made of pixel-strip (PS) detectors, while the rest of the layers are made of strip-strip (2S) detectors. The PS detectors are specially designed for the high occupancy, and harder radiation, closer to the beam pipe.\\
\\
The L1 Track Trigger algorithm starts by reading information from the pair of modules in a single layer, known as a stub, giving a measurement of the transverse momentum (\pt) of a charged particle. Charged particles producing tracks with a large bend (corresponding to a \pt below 2\GeV) are rejected. Next, a tracklet is formed using pairs of neighboring stubs (Fig.~\ref{fig:algorithm} a). These tracklets are used as seeds for extrapolating to other layers, known as projections (Fig.~\ref{fig:algorithm} b). If a projection overlaps with a physical hit in the subsequent layers, a potential track is found. A Kalman filter takes all potential tracks from all seeding combinations, and produces a best track fit (Fig.~\ref{fig:algorithm} c).

\section{Implementation}
The designed implementation of the L1 Track Trigger will run at 240\MHz with a time multiplexing (TMUX) of 18. This TMUX gives each module 18 bx, or 108 clock cycles, to processes data. The entire $\varphi$-region of the detector is divided into nine non-overlapping regions, known as nonants, resulting in at most one sector-to-sector crossing for the 2\GeV minimum track \pt requirement. For the physical hardware, we have chosen to use a field-programmable gate array (FPGA), specifically the Xilinx Virtex UltraScale+\textsuperscript{TM}. The benefits of using FPGAs over application-specific integrated circuits (ASICs) includes: flexible designing, ease of changes and/or updates, off-the-shelf components, and the designs can improve as the industry improves. The track finding algorithm is broken into several modules: Input Router, VM Router, Tracklet Engine, Tracklet Calculator, Projection Router, Match Engine, Match Calculator, Purge Duplicates, and Kalman Filter. The functionality of each is detailed below.

\subsection*{Input Router}
The Input Router (IR) is responsible for collecting all inputs from the DTC, and organizing them into the appropriate $\varphi$~nonants.
\subsection*{VM Router}
The VM Router (VMR) futher divides the outputs of the IR into the various virtual modules (VMs). The VMs allow for fast, parallel processing of stubs in different regions of the detector.
\subsection*{Tracklet Engine}
The Tracklet Engine (TE) is responsible for organizing pairs of adjacent stubs into the seeds required for the track finding algorithm.
\subsection*{Tracklet Calculator}
The Tracklet Calculator (TC) subsequently produces the projections based on the seeds created in the TE.
\subsection*{Projection Router}
The Projection Router (PR) organizes the resulting projections from the TC.
\subsection*{Match Engine}
The Match Engine (ME) is responsible for grouping all possible stub and projection matches created by the TE and TC.
\subsection*{Match Calculator}
The Match Calculator (MC) sorts through the stub/projection pairs from the ME, and finds the best match among the redundant collection. This collection is known as a full match.
\subsection*{Purge Duplicates}
The Purge Duplicates (PD) stage looks for any duplicate full matches the MC may have found, resulting from seeding in multiple layers simultaneously.
\subsection*{Kalman Filter}
The Kalman Filter (KF) takes the final, cleaned, collection from the PD and produces a best-fit track across all layers containing a full match.\\
\\
We are also in the processes of creating super-modules to help with resource management. These modules are known as the Tracklet Processor (TP) and Match Processor (MP). The TP combines the TE and TC stages into a single super-module, while the MP combines the PR, ME, and MC stages into a single super-module.\\
\\
The L1 Track Trigger algorithm was also implemented in a software emulation to measure performance. Using simulated events with top quark pair production, and a PU of 200, an efficiency above 95\% is observed for the desired pseudorapidity coverage ($|{\eta}|<2.5$), as well as a very good resolution on the reconstructed lateral impact parameter, $z_0$.

\section{Hardware}
\begin{figure}[!htb]
    \centering
    \includegraphics[width=0.5\textwidth]{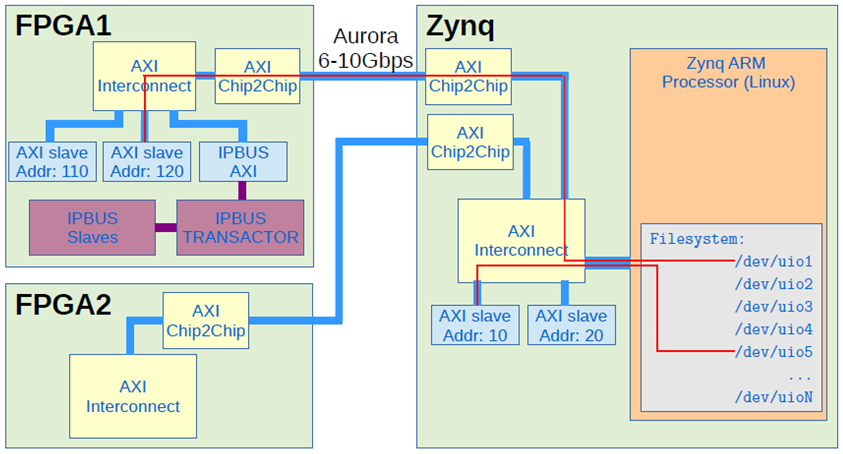}
    \caption{A block diagram of the Apollo module, including the CM (left) and the SM (right).}
    \label{fig:apollo}
\end{figure}

\begin{figure}[!htb]
    \centering
    \includegraphics[width=0.33\textwidth]{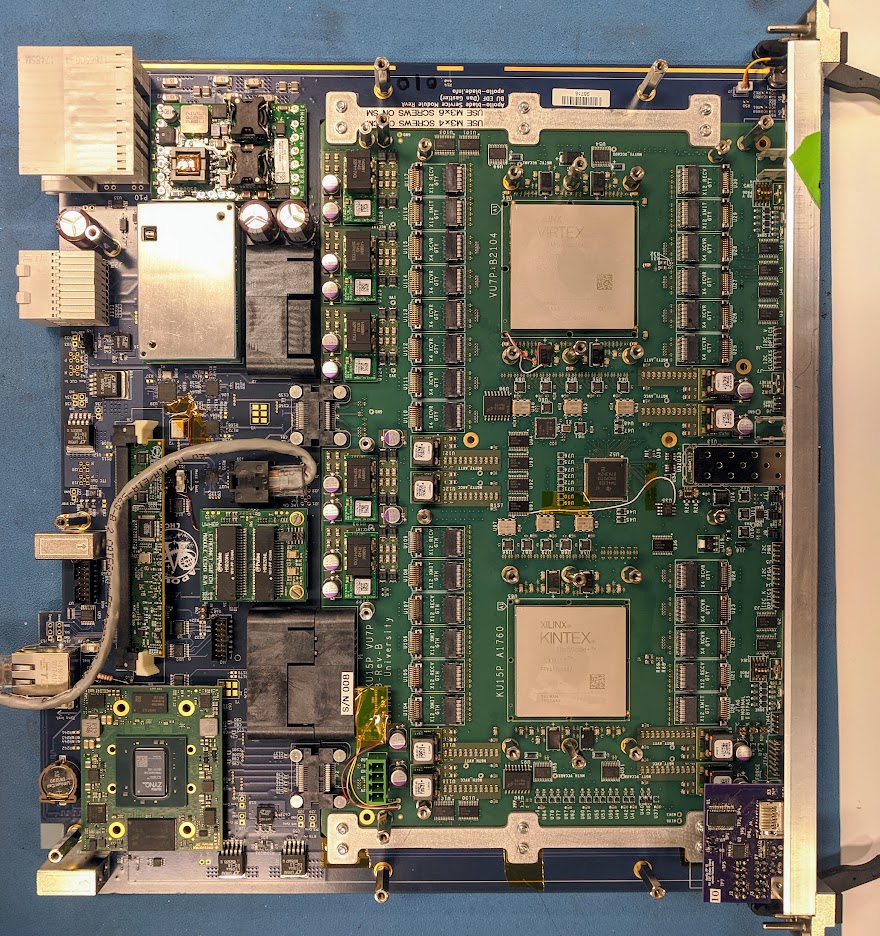}
    \includegraphics[width=0.33\textwidth]{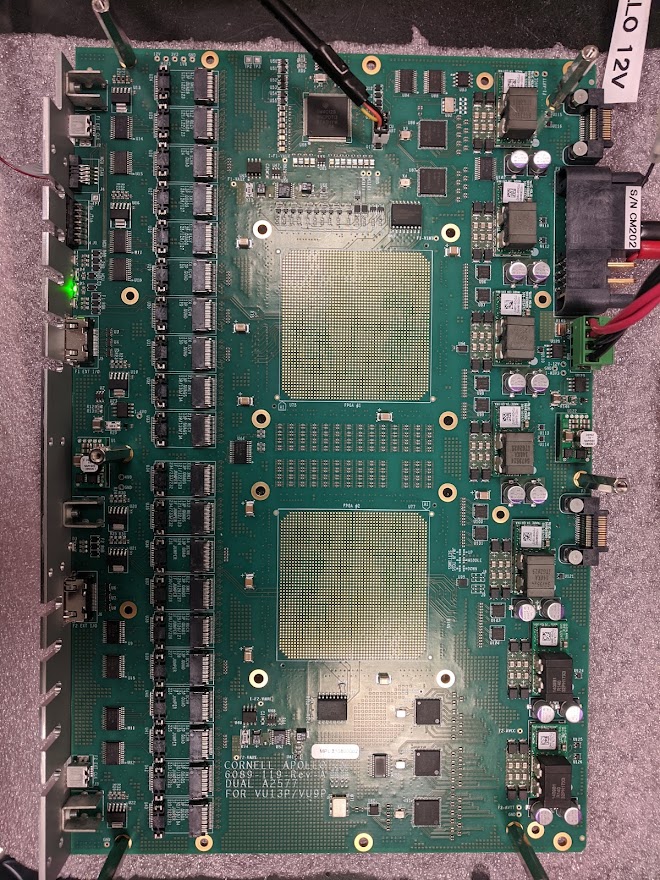}
    \caption{Revision 1 of the SM+CM (left) and revision 2 of the CM (right)}
    \label{fig:rev}
\end{figure}

As mentioned, the target hardware is the Xilinx Virtex UltraScale+\textsuperscript{TM} FPGA chipset. The current demonstrator is using the Apollo module~\cite{Hazen:2020mR}---named after the Apollo missions---which consists of a Service Module (SM) and a Command Module (CM) as shown in Fig.~\ref{fig:apollo}. The SM contains a Zynq system on a chip (SoC) which allows the module to communicate with the outside world. The CM plugs directly into the SM, and contains both a Virtex and Kintex FPGA chip. Demonstrator boards are located at CERN in the CMS Tracker Integration Facility (TIF), Boston University, and Cornell University.
\\
\\
The Apollo boards are currently on revision 1 (Fig.~\ref{fig:rev} left), which has been fully evaluated at a link speed of 25\Gbps via a FireFly optical connector, with a bit error rate below $10^{-16}$. Cornell University is in the processes of testing revision 2 of the CM. Revision 2 (Fig.~\ref{fig:rev} right) of the SM is expected around the end of November 2021. The target delivery date to CERN is mid to late 2024, while installation and commissioning will take place in 2025, and the L1 Track Trigger will be ready for HL-LHC data starting in 2027.

\section{Conclusion}
A robust algorithm for the CMS L1 Track Trigger has been developed to handle the run conditions of the HL-LHS upgrade; it is currently being tested on physical hardware. The algorithm was designed to run at 240\MHz with a time multiplexing of 18. The target hardware is the Xilinx Virtex UltraScale+\textsuperscript{TM}, and the current demonstrator is the Apollo module revision 1. The development of the L1 Track Trigger is on target for delivery, installation, and commissioning between 2024 and 2025. The system will be fully ready for the start of the HL-LHC in 2027.

\bibliography{main}
\bibliographystyle{JHEP}

\end{document}